\documentclass[conference]{IEEEtran}
\IEEEoverridecommandlockouts

\usepackage{cite}
\usepackage{amsmath,amssymb,amsfonts}
\usepackage{algorithmic}
\usepackage{graphicx}
\usepackage{textcomp}
\usepackage{xcolor}

\usepackage{booktabs}               
\usepackage{MnSymbol}               
\usepackage{multirow}
\usepackage{multicol}
\usepackage{svg}
\usepackage{caption} 
\usepackage{subfig}  
\def\BibTeX{{\rm B\kern-.05em{\sc i\kern-.025em b}\kern-.08em
    T\kern-.1667em\lower.7ex\hbox{E}\kern-.125emX}}
\begin{document}

\title{Fast Word Error Rate Estimation \\Using Self-Supervised Representations \\for Speech and Text
\thanks{This work was conducted at the Voicebase/Liveperson Centre of Speech and Language Technology at the University of Sheffield which is supported by Liveperson, Inc..}
}


\author{
\IEEEauthorblockN{Chanho Park, Chengsong Lu\textsuperscript{\textdagger}, Mingjie Chen, Thomas Hain}
\IEEEauthorblockA{\textit{School of Computer Science, University of Sheffield} \\
\textit{Speech and Hearing Research Group}\\
Sheffield, UK \\
\{cpark12, clu22, mingjie.chen, t.hain\}@sheffield.ac.uk}
\thanks{\textdagger Work was done when Chengsong was an intern at the Voicebase/LivePerson Centre.}
}

\maketitle

\begin{abstract}
Word error rate (WER) estimation aims to evaluate the quality of an automatic speech recognition (ASR) system's output without requiring ground-truth labels. This task has gained increasing attention as advanced ASR systems are trained on large amounts of data. In this context, the computational efficiency of a WER estimator becomes essential in practice. However, previous works have not prioritised this aspect. In this paper, a Fast estimator for WER (Fe-WER) is introduced, utilizing average pooling over self-supervised learning representations for speech and text. Our results demonstrate that Fe-WER outperformed a baseline relatively by 14.10\% in root mean square error and 1.22\% in Pearson correlation coefficient on Ted-Lium3. Moreover, a comparative analysis of the distributions of target WER and WER estimates was conducted, including an examination of the average values per speaker. Lastly, the inference speed was approximately 3.4 times faster in the real-time factor.
\end{abstract}

\begin{IEEEkeywords}
Word error rate, WER estimation, self-supervised representation, multi-layer perceptrons, inference speed.
\end{IEEEkeywords}

\begin{figure*}[htbp]
    \centering
    \includegraphics[width=0.95\textwidth]{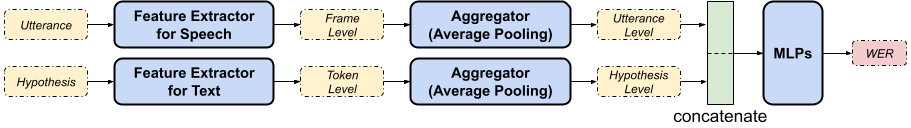}
    \caption{Illustration of the proposed method for WER estimation}
    \label{fig:Illustration of the proposed method for WER estimation}
\end{figure*}

\section{Introduction}
\label{sec:Introduction}
Word error rate (WER) is a commonly used metric for evaluating automatic speech recognition (ASR) systems. It is the ratio of the number of substitution, insertion, and deletion errors in an ASR system's output (hereafter referred to as a hypothesis) to the number of words in a reference. In certain scenarios, it can be beneficial to use a model to estimate the WER of a hypothesis, especially when a ground-truth transcript is unavailable. For example, a WER estimation model can be used to rank hypotheses~\cite{jalalvand-etal-2015-driving} and to select unlabelled data for self-training in ASR~\cite{chen20m_interspeech, 9054295, 9414641}. Another use may be to filter out training data with high-WER transcripts to enhance ASR performance, particularly when collected from the internet. Such data samples are typically excluded from ASR training, especially for recent models like Whisper~\cite{10.5555/3618408.3619590}, which are trained on large datasets sourced online. When dealing with large amounts of data, the computational efficiency of a WER estimator becomes important. One obvious solution to estimate the WER of a hypothesis is to produce confidence scores from the ASR system itself~\cite{kumar20f_interspeech, 9053287}. This method does not require building another model for WER estimation. However, this has the risk of bias and\textemdash as will be shown\textemdash does not perform as well as other WER estimation methods. Moreover, it is poorly aligned with WER due to the lack of prediction of deletion errors. 

Recently, researchers have proposed methods to directly estimate the WER of a hypothesis without the need for ASR decoding. For example, e-WER3~\cite{10095888} used bidirectional long short-term memory (BiLSTM) networks to aggregate speech features, while the text features were averaged over tokens. Next, WER was directly estimated using multi-layer perceptrons (MLPs) with these features. Although it has made impressive progress in estimating the WER of the ASR system's output, there are still several questions that have not been fully studied. Firstly, the e-WER3 model, though avoiding ASR decoding, relies on BiLSTM, which are computationally intensive for long sequences like spoken utterances. This limits their use in training with long speech. Secondly, the performance of the estimator depends on the input features for speech and text. Thus, different combinations of self-supervised learning representations (SSLRs) for speech and text need to be explored for optimal performance on the WER estimation task. Lastly, performance needs to be analysed across data attributes, such as utterance lengths and speakers in addition to the evaluation metrics. 

In this paper, a fast estimator for WER (Fe-WER) is proposed, utilising SSLRs aggregated through average pooling. The model comprises speech and text encoders, feature aggregators, and an estimator to directly predict WER using the aggregated representations. This approach is examined from both accuracy and efficiency perspectives. The contributions of this paper are as follows:
\begin{enumerate}
    \item This paper proposes Fe-WER, a WER estimation model that uses average pooling and outperforms the baseline model in computational efficiency without compromising performance in WER estimation.
    \item Experimental evidence shows that the combination of HuBERT \cite{9585401} and XLM-R \cite{conneau-etal-2020-unsupervised} achieves the best performance in WER estimation.
    \item A comparative analysis of the distributions of target WER and WER estimates is presented including an examination of the average values per speaker.
\end{enumerate}

\section{Related works}
\label{sec:Related works}
\subsection{WER Estimation}
\label{subsec:WER estimation}
e-WER3 \cite{10095888} is a WER estimator for multiple languages. For generating training data, hypotheses for Ted-Lium3 \cite{10.1007/978-3-319-99579-3_21} were generated using an ASR system. Utterances and transcripts were encoded using XLSR-53 \cite{conneau21_interspeech} and XLM-R \cite{conneau-etal-2020-unsupervised}, respectively. The hidden states of BiLSTM in both directions over frame-level representations were concatenated to form an utterance-level representation, while a transcript-level representation was averaged over token-level representations. To address data imbalance, hypotheses with a WER of 0 were chosen, up to the total count of entries in the second and third most frequent histogram bins (out of 100). The WER was predicted using MLPs on top of the concatenated representation. The result was 0.14 in root mean square error (RMSE) and 0.66 in Pearson correlation coefficient (PCC), which was improved relatively by 9\% in PCC from e-WER2~\cite{ali20_interspeech}.

\subsection{Sequence-Level Representation}
\label{subsec:Sequence representation}
In \cite{reimers-gurevych-2019-sentence}, a sentence-level representation was suggested for NLP tasks, such as semantic textual similarity between sentences. The representation, called SBERT, was learned using a Siamese or a triplet model\textemdash often referred to as a two-tower architecture~\cite{10.1145/2505515.2505665, 10.1145/3366424.3386195}\textemdash with classification, regression and triplet objective functions. BERT~\cite{devlin2018bert}, one of the SSLRs, was adopted and converted into a fixed-length representation for a sentence through different pooling strategies. The results showed that the average pooling strategy outperformed the others, such as using a special token for classification of BERT. In addition to SBERT, the average pooling strategy for utterance-level representation has gained popularity in many other tasks, such as speaker identification, intent classification and emotion recognition~\cite{wang-etal-2018-glue, yang21c_interspeech}. 

\section{Fast Word Error Rate Estimation}
\label{sec:Fast word error rate estimator}
\subsection{Architecture}
Fe-WER (see Fig.~\ref{fig:Illustration of the proposed method for WER estimation}) is based on a two-tower architecture that maps different representations into a shared space. The proposed model consists of two aggregators\textemdash one for speech and another for text\textemdash and MLPs that perform the WER estimation. The aggregators convert the features extracted by SSLRs into sequence-level representations. These two sequence-level representations are concatenated and input to MLPs consisting of fully connected layers with a rectified linear unit (ReLU) activation function. A sigmoid function is applied to the output. The WER estimate is defined:
\begin{equation*}
\label{equation:WER prediction}
    \widehat{\text{WER}} = \text{MLP}(\text{concat}(a(f(s)), a(g(t))))
\end{equation*}
where $a$ is a function of average pooling, $f(\cdot)$ and $g(\cdot)$ are speech and text encoders, respectively, and $s$ and $t$ are a spoken utterance and its corresponding hypothesis, respectively.

\subsection{Training Objective}
The mean squared error (MSE) between WER and $\widehat{\text{WER}}$ is used as the objective function to train the MLPs, where WER represents the error rate between a reference and a hypothesis and $\widehat{\text{WER}}$ is the estimation by the model.
\begin{equation*}
    \text{MSE} = \frac{\sum_{i=1}^N(\text{WER}_{i} - \widehat{\text{WER}}_{i})^2}{N}
\end{equation*}
where $N$ is the number of instances in a dataset and $i$ is the index of an instance. 

\subsection{Weighted Word Error Rate Estimate}
The WER can be weighted by the number of words in a reference transcript, denoted as $\text{WER}_{\text{wrd}}$. For the weighted WER estimation on a dataset without reference transcripts, it is weighted by duration instead of the number of words in the references. The weighted WER estimate is defined as follows:
\begin{equation*} \label{eq:weighted WER by duration}
    \widehat{\text{WER}}_{\text{dur}} = \frac{\sum_{i=1}^N(\widehat{\text{WER}}_{i} \times \text{Duration}_{i})}{\sum_{i=1}^N(\text{Duration}_{i})}
\end{equation*}
where $i$ is the index of a pair consisting of an utterance and its corresponding hypothesis. 

\begin{table*}[htbp]
\caption{Statistics of the selected data sets. Hypotheses were generated using Whisper large-v2.}
\label{table:Statistics of the sets of data selected}
    \begin{center}
    \begin{tabular}{ c|c|c|c|c|c|c|c }
    \toprule
    Dataset & \#Seg. & Total Dur. (h) & Avg. Dur. & Avg. \#Wrd. & Avg. WER & Std. Dev. of WER & WER$_{\text{wrd}}$ \\
    \hline
    \hline
    test  & 842    & 1.41    & 6.05 & 16.72 & 14.29\% & 19.97\% & 10.88\% \\
    dev   & 1034   & 1.70    & 5.93 & 17.72 & 15.32\% & 22.47\% & 12.25\% \\
    train & 123255 & 200.55  & 5.86 & 17.04 & 24.34\% & 32.09\% & 20.29\% \\
    \bottomrule
    \end{tabular}
    \end{center}
\end{table*}

\subsection{Evaluation Metrics}
RMSE and PCC are used as evaluation metrics. RMSE is the root of MSE, while PCC measures linear association, ranging from -1 (negative) to +1 (positive), with 0 indicating no correlation.
\begin{equation*}
    \frac{\sum_{i=1}^N(\text{WER}_{i}-\mu_{\text{WER}})(\widehat{\text{WER}}_{i}-\mu_{\widehat{\text{WER}}})}{\sqrt{\sum_{i=1}^N(\text{WER}_{i}-\mu_{\text{WER}})^2 \sum_{i=1}^N(\widehat{\text{WER}}_{i}-\mu_{\widehat{\text{WER}}})^2}}
\end{equation*}
where $\mu_{\text{WER}}$ is the mean of WER. For weighted WER estimation, the ratio between the weighted $\text{WER}_{\text{wrd}}$ and $\widehat{\text{WER}}_{\text{dur}}$ (WERR) is also measured.
\begin{equation*}
    \text{WERR} = \frac{|\text{WER}_{\text{wrd}} - \widehat{\text{WER}}_{\text{dur}}|}{\text{WER}_{\text{wrd}}}.
\end{equation*}

\section{Experiment Setup}
\subsection{Data} \label{subsec:Data} 
Ted-Lium3 (TL3)~\cite{10.1007/978-3-319-99579-3_21} was used for WER estimation. Whisper large-v2~\footnote{https://github.com/openai/whisper} was employed to transcribe the corpus due to its comparable performance on TL3, reproducibility and public availability. Whisper’s text normaliser was employed after being modified to prevent the replacement of numeric expressions with Arabic numerals. After the text normalisation, the data imbalance due to the high volume of WER 0 was addressed as described in Section~\ref{subsec:WER estimation}. For comparison with baseline systems, utterances with lengths up to 10 seconds were selected, and WER was clamped between 0\% and 100\%. The statistics of the selected data are summarised in Table \ref{table:Statistics of the sets of data selected}. The training set’s higher WER might be due to additional data in TL3 introducing varied conditions, while the dev and test sets remain unchanged from the previous version.

\subsection{Self-Supervised Learning Representations}
\label{subsec:Self-supervised learning representations}
SSLRs for utterances and hypotheses were selected based on their performance on benchmarks including Speech processing Universal PERformance Benchmark (SUPERB) \cite{yang21c_interspeech}, General Language Understanding Evaluation (GLUE) \cite{wang-etal-2018-glue} and SuperGLUE \cite{NEURIPS2019_4496bf24}. These benchmarks assess models on various tasks, such as phoneme recognition and paraphrase detection. Additionally, two SSLR models used in \cite{10095888} for WER estimation were included for comparison. Summary information on these models, including model size and the number of parameters, is provided in Table \ref{table:models summary}. 
\begin{table}[htbp]
\caption{Summary information of SSLRs.}
\label{table:models summary}
    \begin{center}
    \begin{tabular}{ c | l c c c }
     \hline
     Input Type & \multicolumn{1}{c}{Model} & Abbr. & Size & \#Parameters \\
     \hline
     \hline
     \multirow{4}{*}{Utterance} & data2vec~\cite{baevski2022data2vec} & DAT & Large & 313M \\
      & HuBERT~\cite{9585401} & HUB & Large & 316M \\
      & WavLM~\cite{chen2022wavlm} & WAV & Large & 317M \\
      & XLSR-53~\cite{conneau21_interspeech} & XLS & Large & 315M \\
     \hline
     \multirow{4}{*}{Transcript} & DeBERTa-V3~\cite{he2021debertav3} & DEB & Large & 283M \\
      & GPT-2~\cite{radford2019language} & GPT & Medium & 355M \\
      & RoBERTa~\cite{liu2020roberta} & ROB & Large & 355M \\
      & XLM-R~\cite{conneau-etal-2020-unsupervised} & XLM & Large & 560M \\
     \hline
    \end{tabular}
    \end{center}
\end{table}

\subsection{Baseline WER Estimators}
\label{subsec:Baseline WER estimators} 
The proposed model was compared with two baselines: a method using a confidence score (WER-CS) and another with BiLSTM. First, for sequence-level confidence scoring, the log probability of Whisper large-v2 over the output tokens was averaged and subtracted from 1. For decoding, two strategies were employed: greedy decoding only and full decoding. The full decoding strategy included a beam size of 5, greedy decoding with the 5 best hypotheses and sampling temperature settings ranging from 0 to 1 in increments of 0.2. Second, a WER estimation model employed BiLSTM for aggregation. Single-layer BiLSTM networks were used to aggregate frame-level SSLR representations, with the hidden feature size matching that of the input features. For further details, readers can refer to e-WER3~\cite{10095888}.

\subsection{Fe-WER}
\label{subsec:Fe-WER} 
Average pooling over the frame or token dimension was used as an aggregator. A Fe-WER model includes MLPs with two hidden layers and an output layer, activated by ReLU and Sigmoid functions, respectively. The layers consist of 600, 32, and 1 nodes on top of 2048-dimensional input features. Each layer's output is normalised except for the output layer, and dropout (0.1) is applied to the hidden layers. The model was trained with an Adam optimiser (learning rate: 1e-3), a cosine annealing scheduler (max iterations: 15) and early stopping at 40 epochs. Hyperparameters were selected via grid search. 


\section{Results}
\label{sec:Results}
Aggregators were compared across various SSLR combinations, followed by WER model comparisons with confidence scoring baselines, utterance-level analysis, and inference speed evaluation.

\subsection{Aggregators} \label{subsec:Aggregators} 
BiLSTM and average pooling are compared using combinations of SSLRs in Section \ref{subsec:Self-supervised learning representations}. First, RMSE and PCC tend to improve with average pooling in 13 out of 16 combinations. Second, the best combinations are DAT and XLM for BiLSTM and HUB and XLM for average pooling. The latter outperformed the former by 0.0099 in RMSE and 0.0228 in PCC on TL3 dev. Results are summarised in Table~\ref{table:Results of BiLSTM and Average pooling aggregators with different SSLRs and three seeds on TL3 dev}.
\setlength\tabcolsep{5pt}
\begin{table}[htbp]
    \caption{Results of BiLSTM and Average pooling aggregators with different SSLRs and three seeds on TL3 dev.}
    \label{table:Results of BiLSTM and Average pooling aggregators with different SSLRs and three seeds on TL3 dev}
    \begin{center}
    \begin{tabular}{ c|c|c|c|c|c }
    \toprule
    \multicolumn{2}{c|}{SSLR} & \multicolumn{2}{c|}{BiLSTM}       & \multicolumn{2}{c}{Average Pooling}   \\
    \hline                                                                                                  
    Utt.      & Hyp.          & RMSE$\downarrow$ & PCC$\uparrow$    & RMSE$\downarrow$ & PCC$\uparrow$    \\
    \hline                                                                                               
    \hline                                                                                               
    DAT        & DEB            & .1185$\pm$.001 & .8490$\pm$.004 & .1213$\pm$.000 & .8425$\pm$.001 \\
    DAT        & GPT            & .1254$\pm$.005 & .8405$\pm$.008 & .1185$\pm$.001 & .8512$\pm$.002 \\
    DAT        & ROB            & .1193$\pm$.002 & .8491$\pm$.008 & .1190$\pm$.002 & .8486$\pm$.004 \\
    DAT        & XLM            & \textbf{.1111$\pm$.008} & \textbf{.8700$\pm$.018} & .1137$\pm$.001 & .8637$\pm$.002 \\
    \hline                                                                                                  
    HUB        & DEB            & .1216$\pm$.002 & .8398$\pm$.004 & .1105$\pm$.002 & .8702$\pm$.005 \\
    HUB        & GPT            & .1233$\pm$.002 & .8387$\pm$.005 & .1093$\pm$.001 & .8741$\pm$.001 \\
    HUB        & ROB            & .1227$\pm$.004 & .8363$\pm$.011 & .1123$\pm$.003 & .8676$\pm$.006 \\
    HUB        & XLM            & .1212$\pm$.011 & .8418$\pm$.032 & \textbf{.1012$\pm$.003} & \textbf{.8928$\pm$.007} \\
    \hline                                                                                                  
    WAV        & DEB            & .1289$\pm$.005 & .8200$\pm$.014 & .1164$\pm$.002 & .8551$\pm$.003 \\
    WAV        & GPT            & .1270$\pm$.003 & .8245$\pm$.009 & .1111$\pm$.002 & .8709$\pm$.006 \\
    WAV        & ROB            & .1210$\pm$.004 & .8420$\pm$.013 & .1167$\pm$.002 & .8561$\pm$.004 \\
    WAV        & XLM            & .1172$\pm$.005 & .8520$\pm$.015 & .1099$\pm$.002 & .8734$\pm$.005 \\
    \hline                                                                                                  
    XLS        & DEB            & .1289$\pm$.003 & .8191$\pm$.011 & .1216$\pm$.002 & .8412$\pm$.006 \\
    XLS        & GPT            & .1200$\pm$.003 & .8467$\pm$.008 & .1155$\pm$.001 & .8585$\pm$.002 \\
    XLS        & ROB            & .1285$\pm$.003 & .8226$\pm$.006 & .1161$\pm$.003 & .8567$\pm$.007 \\
    XLS        & XLM            & .1199$\pm$.005 & .8474$\pm$.009 & .1101$\pm$.001 & .8717$\pm$.003 \\
    \bottomrule
    \end{tabular}
    \end{center}
\end{table}

\subsection{Comparison with Baselines} 
The proposed model, which uses an average pooling aggregator with HUB and XLM, is compared to WER-CS and a model using BiLSTM with DAT and XLM. First, WER-CS with the two decoding strategies described in Section~\ref{subsec:Baseline WER estimators} performed worse than the other models in both metrics, while the proposed model outperformed the BiLSTM baseline with relative improvements of 14.10\% in RMSE and 1.22\% in PCC. Second, in terms of WERR, models using SSLRs estimate the WER of a test set within 5\% of the target, while WER-CS models overestimate it by more than double. The comparison results are shown in Table~\ref{table:RMSE and PCC of baseline systems on the TL3 test}. 
\begin{table}[htbp]
\caption{RMSE and PCC of baseline systems on TL3 test. $\text{WER}_{\text{wrd}}$ is a target WER weighted by words. $\widehat{\text{WER}_{\text{dur}}}$ is the WER estimate weighted  by duration. \textdagger\ is the proposed method.}
\label{table:RMSE and PCC of baseline systems on the TL3 test}
    \begin{center}
    \begin{tabular}{ l l c c c c }
     \toprule
      & RMSE$\downarrow$ & PCC$\uparrow$ & $\text{WER}_{\text{wrd}}$ & $\widehat{\text{WER}_{\text{dur}}}$ & WERR$\downarrow$ \\
     \hline
     \hline
     WER-CS      &       &       &        &        &       \\
     + full      & 0.2611 & 0.5654 & 8.40\%  & 31.85\% & 279.16\% \\
     + greedy    & 0.2546 & 0.6944 & 10.88\% & 33.34\% & 206.43\% \\
     \hline
     BiLSTM      &       &       &        &        &       \\
     + DAT,XLM     & 0.1071 & 0.8793 & 10.88\% & 10.96\% & 0.73\% \\
     \hline
     \textdagger Avg. Pool.  &       &       &        &        &       \\
     + HUB,XLM     & 0.0920 & 0.8900 & 10.88\% & 10.39\% & 4.50\% \\ 
     \bottomrule
    \end{tabular}
    \end{center}
\end{table}

\subsection{Distributions of Target WER and WER Estimates} \label{subsec:Distributions of target WER and WER estimate} 
\begin{figure}[htbp]
     \centering
     \subfloat[][target WER]{\includegraphics[width=4.2cm]{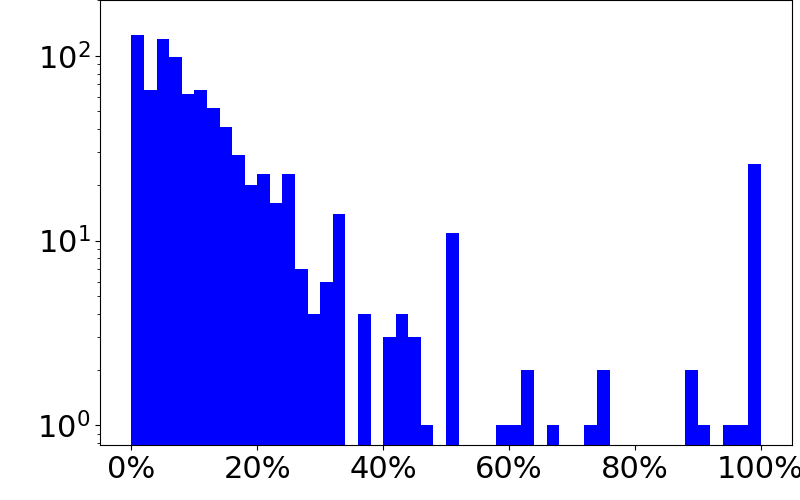} \label{fig:target WER}}
     \subfloat[][WER estimate]{\includegraphics[width=4.2cm]{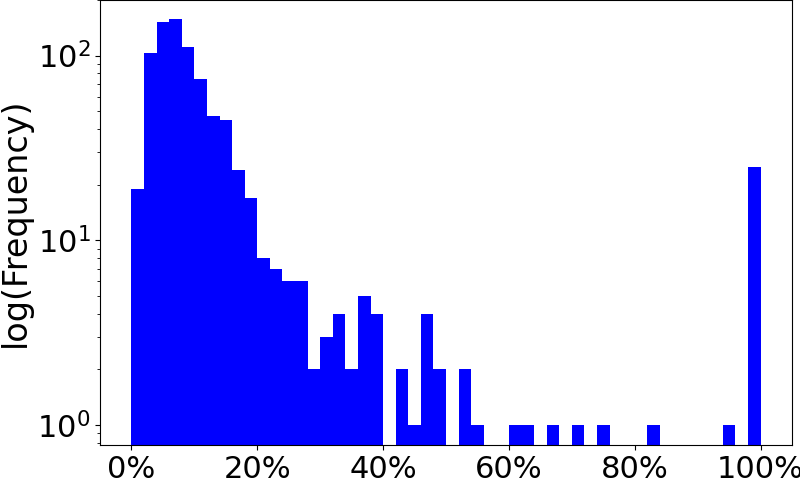} \label{fig:WER estimate}}
\caption{Histograms on TL3 test} \label{fig:Histograms of target WER and estimates on TL3 test}
\end{figure}

The histograms of target WERs and WER estimates on TL3 test are visualised in Fig.~\ref{fig:Histograms of target WER and estimates on TL3 test}. The distribution of Fe-WER estimates is similar to that of the target WERs. However, the frequency of target WERs peaks between 0 and 2 percent (exclusive of 2) in Fig.~\subref{fig:target WER}, while the estimates peak between 4 and 8 percent (exclusive of 8) in Fig.~\subref{fig:WER estimate}. This discrepancy may be due to the Sigmoid function outputting small values rather than 0. Additionally, WER estimates above 20\% are generally less frequent than target WERs. In this range, three or more insertions in a row are frequently observed in the hypotheses. Therefore, recognising these words as one insertion error may have led to the low estimates. 

\subsection{Average Target WER and WER Estimate per Speaker} \label{subsec:Average of target WER and WER estimates per speaker} 
\begin{figure}[htb]
  \centering
  \includegraphics[width=0.90\linewidth]{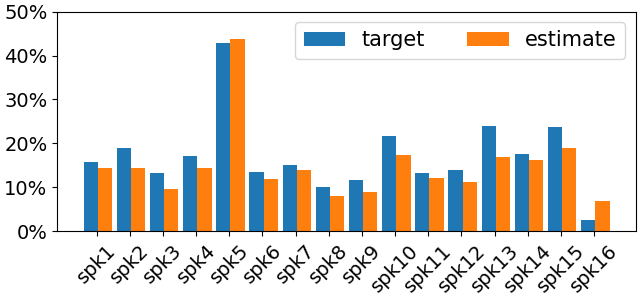}
  \caption{Average WER per each speaker}
  \label{fig:Average WER of each speaker}
\end{figure}

The distributions of average target WER and WER estimate per speaker are similar (see Fig.~\ref{fig:Average WER of each speaker}). The high average target WER of Speaker 5 is due to the majority of shorter utterances, which have low resolution of WER. For example, the WER of a spoken utterance for a word is 0 or at least 100\%. For Speaker 16, the average WER estimate is higher than the average WER target due to the low WER. The phenomenon of high WER estimate was discussed in Section~\ref{subsec:Distributions of target WER and WER estimate}. 

\subsection{Inference Speed} 
The inference time of the WER estimators was measured on a single NVIDIA RTX A6000 GPU with a batch size of 1, including encoding time. The baseline model using BiLSTM had an inference time of 18.64 seconds, while the proposed method's inference time was significantly shorter at 5.42 seconds, reducing the inference time by approximately 70.92\%. The details are summarised in Table \ref{table:Inference time and real-time factor (RTF) of BiLSTM and Average pooling estimators and using Whisper large-v2}. 
\begin{table}[htbp]
\caption{Inference time (in seconds) and real-time factor (RTF) of BiLSTM and Avg. Pool. with HUB and XLM on TL3 test. Total duration is approximately 5223 seconds. RTF: total time $\div$ total duration. \textdagger\ is the proposed method.}
\label{table:Inference time and real-time factor (RTF) of BiLSTM and Average pooling estimators and using Whisper large-v2}
    \begin{center}
    \begin{tabular}{ lcc }
    \toprule
    & BiLSTM & \textdagger Avg. Pool. \\
    \hline
    \hline
    Feature extraction &       &            \\
    + utterance        & \multicolumn{2}{c}{2.72} \\
    + transcript       & \multicolumn{2}{c}{0.93} \\
    Aggregation        & 5.28  & $\epsilon$ \\
    Feedforward        & 9.71  & 1.77       \\
    \hline
    Total              & 18.64 & 5.42       \\
    \hline
    RTF                & 0.003569 & 0.001038   \\
    \bottomrule
    \end{tabular}
    \end{center}
\end{table}

\section{Conclusion}
\label{sec:Conclusion}
In this paper, a fast WER estimator is proposed. The proposed model consists of speech and text encoders for SSLRs, aggregators using average pooling and an MLP estimator. The WER estimator outperforms the BiLSTM baseline by relative 14.10\% and 1.22\% in RMSE and PCC, respectively. Moreover, the experimental results show that the inference speed has been significantly improved, being 3.4 times faster than the BiLSTM baseline, without performance degradation. 







\bibliographystyle{plain} 
\bibliography{refs} 

\end{document}